\documentclass[conference]{IEEEtran}

\usepackage{cite}
\usepackage{graphicx,color,epsfig,rotating}
\usepackage{amsfonts,amsmath,amssymb,bbm}
\usepackage{algorithm,algorithmic}
\usepackage{subfigure}
\usepackage{amsmath}
\usepackage{cite}
\usepackage{mdwtab}
\usepackage{placeins}
\usepackage{psfrag, graphicx}
\usepackage[latin1]{inputenc}
\usepackage{amssymb}
\usepackage{multirow}

\long\def\comment#1{}

\newfont{\bbb}{msbm10 scaled 700}

\newfont{\bb}{msbm10 scaled 1100}

\newcommand{\PP}{\mbox{\bb P}}

\newcommand{\FF}{\mbox{\bb F}}

\newcommand{\Ac}{{\cal A}}

\newcommand{\Dc}{{\cal D}}

\newcommand{\Fc}{{\cal F}}

\newcommand{\Lc}{{\cal L}}

\newcommand{\Uc}{{\cal U}}

\newcommand{\fsf}{{\sf f}}

\newcommand{\ssf}{{\sf s}}

\newcommand{\Asf}{{\sf A}}

\newcommand{\be}{\begin{equation}}
\newcommand{\ee}{\end{equation}}
\newcommand{\bea}{\begin{eqnarray}}
\newcommand{\eea}{\end{eqnarray}}

\def\fsf{ {\sf f}}

\newtheorem{defn}{Definition}
\newtheorem{theorem}{Theorem}
\newtheorem{corollary}{Corollary}

\begin{document}

\sloppy

\title{Fundamental Limits of Distributed Caching in D2D Wireless Networks}

\author{
  \IEEEauthorblockN{Mingyue Ji, Giuseppe Caire and Andreas F. Molisch}
  \IEEEauthorblockA{Department of Electrical Engineering\\
    University of Southern California\\
    Email: \{mingyuej, caire, molisch\}@usc.edu} 
}




\maketitle

\begin{abstract}
We consider a wireless Device-to-Device (D2D) network where communication is restricted to be single-hop, users make arbitrary 
requests from a finite library of possible files and user devices cache information in the form of linear combinations of packets from the files in the library (coded caching).
We consider the combined effect of coding in the caching and delivery phases, achieving ``coded multicast gain'', and of spatial reuse due to local short-range 
D2D communication. Somewhat counterintuitively, we show that the coded multicast gain and the spatial reuse gain do not cumulate, in terms of the 
throughput scaling laws. In particular, the spatial reuse gain shown in our previous work on uncoded random caching 
and the coded multicast gain shown in this paper yield the same scaling laws behavior, 
but no further scaling law gain can be achieved by using both coded caching {\em and} D2D spatial reuse.  
\end{abstract}

\section{Introduction}
\label{section: intro}

Wireless traffic is dramatically increasing, mainly due to on-demand video streaming \cite{cisco66}. 
One of the most promising approaches for solving this problem is \emph{caching}, i.e. storing video files 
in the users' local caches and/or in dedicated helper nodes disseminated in the network coverage area \cite{6495773, DBLP:journals/corr/abs-1109-4179, ji2013optimal, gitzenis2012asymptotic}. Capitalizing on the fact that 
user demands are highly redundant (e.g., $n \approx$ 10000 users in a university campus 
streaming movies from a library of $m \approx 100$ popular titles), each user demand can be satisfied through 
local communication from a cache, without requiring a high-throughput backhaul to the core network. Such backhaul would constitute a major bottleneck, being too costly or (in the case of mobile helper nodes) completely infeasible. 
in the case of wireless helper nodes.

In particular,  a one-hop Device-to-Device (D2D) communication network with caching at the user nodes is studied in \cite{ji2013optimal}.
The network is formed by $n$ user nodes, each of which stores $M$ files from a library of $m$ files. 
Under the simple protocol model of \cite{gupta2000capacity}, we showed that by using a well-designed random caching policy 
and interference-avoidance transmission with spatial reuse, such that links sufficiently separated in space can be simultaneously active, as 
$n, m \rightarrow \infty$ with $n \gg m$ the throughput per user behaves as $\Theta\left(\frac{M}{m}\right)$ and the outage probability, i.e., 
the probability that a user request cannot be served, is negligible. Furthermore, this scaling is shown to be order-optimal under 
the considered network model. \footnote{Notation: given two functions $f$ and $g$, we say that: 1)  $f(n) = O\left(g(n)\right)$ if there exists a constant $c$ and integer $N$ such that  $f(n)\leq cg(n)$ for $n>N$. 
2) $f(n) = \Theta\left(g(n)\right)$ if $f(n) = O\left(g(n)\right)$ and $g(n) = O\left(f(n)\right)$.}

A different approach to caching is taken in \cite{maddah2012fundamental}, where a system with a single transmitter (e.g., a cellular base station) serving 
$n$ receivers (users) is considered.  The user caches have again size of $M$ files. 
However, instead of caching individual files or segments thereof, coded caching is used. Files are divided into packets (sub-packetization), 
and carefully designed linear combinations of such packets are cached. 
The delivery phase consists of the multi-cast transmission of a sequence of coded packets
by the base station such that the maximum required number of transmitted packets over any arbitrary set of users' demands is minimized. 
The scheme of \cite{maddah2012fundamental} achieves min-max number of transmissions that is given by
$n\left(1-\frac{M}{m}\right)\frac{1}{1+\frac{Mn}{m}}$. 
This scheme is shown to be approximately optimal, by developing a  cut-set lower bound on the min-max number of transmissions. 
Notice that for $n \gg m$, the throughput scaling is again given by $\Theta\left (\frac{M}{m} \right )$. 

Notice that a conventional system, serving each user demand as an individual TCP/IP connection to some video server 
in some CDN \cite{nygren2010akamai} placed in the core network, as it is currently implemented today, 
yields per-user throughput scaling $\Theta \left (\frac{1}{n} \right )$. Instead, both the caching approaches of \cite{ji2013optimal} and of 
\cite{maddah2012fundamental} yield $\Theta \left (\frac{M}{m} \right )$, which is a much better scaling for $n \gg m$, i.e., 
in the regime of highly redundant demands, for which caching is expected to be efficient. The D2D approach of  \cite{ji2013optimal} makes use
of the spatial reuse of D2D local communication, while the approach of \cite{maddah2012fundamental} makes use of coding.
In terms of throughput scaling laws, D2D spatial reuse and coded caching yield the same gain over a conventional system.
Furthermore, both the achieved spatial reuse and the achieved coded caching gains are shown to be optimal. 

\section{Overview of the Main Results}
\label{section: overview}

A natural question at this point is whether any gain can be obtained by {\em combining} spatial reuse and coded caching. 
In this paper, we consider the same model of D2D wireless network as  \cite{ji2013optimal}, but we consider coded caching and delivery phases. 
The main contributions of this paper are as follows: 1) if no spatial reuse is possible (i.e., only one concurrent transmission is allowed in the network), 
the proposed coded caching and delivery scheme with sub-packetization achieves almost the same throughput of \cite{maddah2012fundamental}, without 
the need of a base station; 2) when spatial reuse is possible, then for any combination of spatial reuse and coded caching, the throughput has the same 
scaling law (with possible different constant) of the reuse-only case  \cite{ji2013optimal} or the coded-only case \cite{maddah2012fundamental}. 
Counterintuitively, this means that
it is not possible to cumulate the spatial reuse gain and the coded caching gain, as far as the throughput scaling law 
is concerned. It follows that the best combination of reuse and coded caching gains must be sought in terms of the actual throughput in bit/s/Hz (i.e., 
in the constants at large, but finite, $n,m,M$), rather than in terms of scaling laws.

The paper is organized as follows. Section~\ref{sec: Network Model and Problem Formulation} presents the network model and the formal 
problem definition.  We illustrate all the main results in Section~\ref{sec: Main Results} and give some discussions in 
Section~\ref{sec: Discussions}. Due to the space limit, proofs and details are omitted and can be found in \cite{mingyue20132}.

\section{Network Model and Problem Definition}
\label{sec: Network Model and Problem Formulation}

\begin{figure}
\centering
\subfigure[]{
\centering \includegraphics[width=3.5cm]{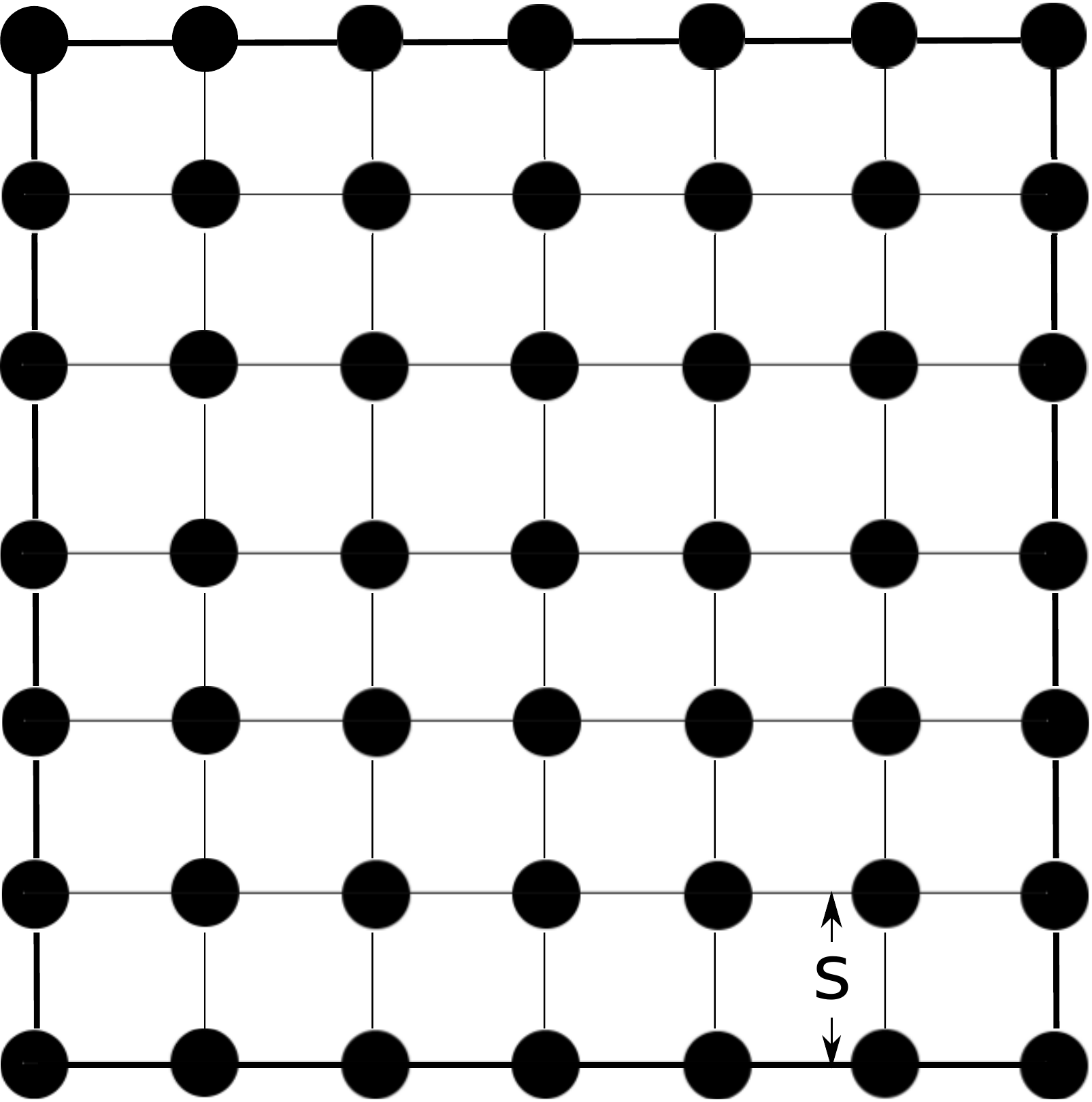}
\label{fig: Grid_Network_D2D}
}
\subfigure[]{
\centering \includegraphics[width=3.5cm]{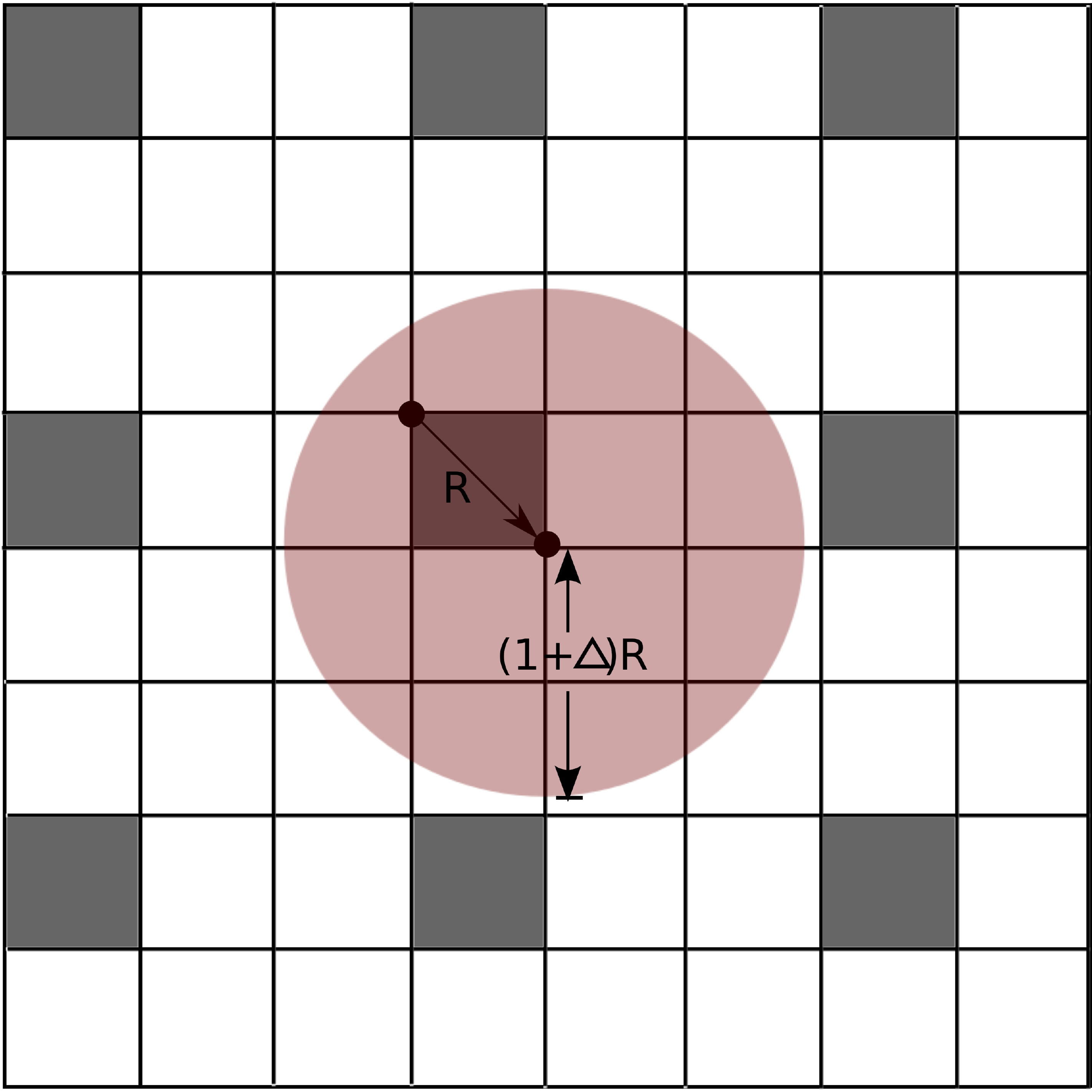}
\label{fig: Grid_TDMA}
}
\caption{a)~Grid network with $n=49$ nodes (black circles) with minimum separation $s = \frac{1}{\sqrt{n}}$. 
b)~An example of single-cell layout and the interference avoidance TDMA scheme. 
In this figure, each square represents a cluster. 
The gray squares represent the concurrent transmitting clusters. 
The red area is the disk where the protocol model allows no other concurrent transmission. 
$R$ is the worst case transmission range and $\Delta$ is the interference parameter. 
We assume a common $R$ for all the transmitter-receiver pairs. 
In this particular example, the TDMA parameter is $K=9$.}
\end{figure}

We consider a grid network formed by $n$ nodes $\Uc = \{u_1, \ldots, u_n\}$ 
placed on a regular grid on the unit square, with minimum distance  $1/\sqrt{n}$. (see Fig.~\ref{fig: Grid_Network_D2D}).
Users $u \in \Uc$ make arbitrary requests $f_u \in \Fc = \{f_1, \ldots, f_m\}$, from a fixed file library of size $m$. 
The vector of requests is denoted by $\fsf = (f_{u_1}, \ldots, f_{u_n})$. 
Communication between user nodes obeys the following {\em protocol model}: if a node $i$ transmits a packet to node $j$, 
then the transmission is successful if and only if: a) The distance between $i$ and $j$ is less than $r$; b) Any other node 
$k$ transmitting simultaneously, is at distance $d(k,j) \geq (1+\Delta) r$ from the receiver $j$, where 
$r, \Delta > 0$ are protocol parameters. 

In practice, nodes send data at some constant rate $C_r$ bit/s/Hz, where $C_r$ is a non-increasing function of the transmission range $r$ \cite{DBLP:journals/corr/abs-1109-4179}.

\begin{figure}
\centering
\includegraphics[width=7cm]{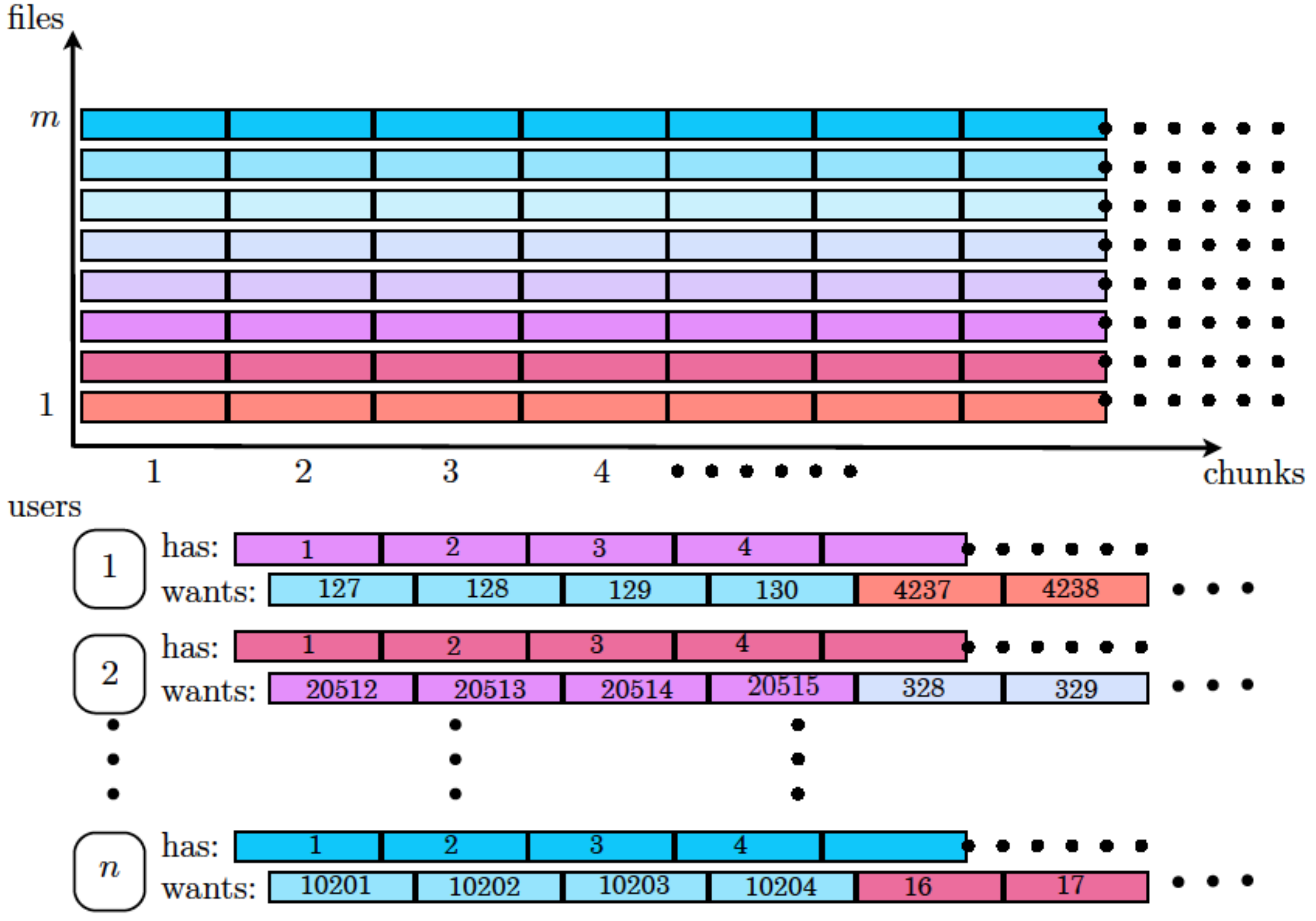} 
\caption{Qualitative representation of our system assumptions: each user caches an entire file, formed by an arbitrarily large number of chunks.
Then, users place random requests of finite sequences of chunks from files of the library, or random duration and random initial points.}
\label{library-and-asynchronous-demands}
\end{figure}

Unlike live streaming,  in video on-demand, the probability that two users wish to stream simultaneously a file at the {\em same time} 
is essentially zero, although there  is a large  redundancy in the demands when $n \gg m$. 
In order to model the intrinsic {\em asynchronism} of video on-demand 
and forbid any form of {\em uncoded multicasting gain} by overhearing ``for free''  transmissions dedicated to other users,  we assume that each file in the 
library is formed by $L$ packets.\footnote{This is compliant with current video streaming protocols such as 
DASH  \cite{6495773}, where the video file is split into segments which are sequentially downloaded by the streaming users.} 
Then, we assume each user downloads a randomly selected segment of length $L'$ packets of the requested file, 
as qualitatively shown in Fig.~\ref{library-and-asynchronous-demands}. 
According to our model, a demand vector $\fsf$ is associated with a list of random pointers 
$\ssf$ with elements $s_u \in \{1, \ldots, L - L'+1\}$ such that for each $u$ demanding file $f_u$
the corresponding segment of packets $s_u, s_u+1, \ldots, s_u+L'-1$ is downloaded. 
Here, $\ssf$ is random i.i.d., and it is known as side information by all nodes, and explicit dependency on $\ssf$ is omitted for simplicity of notation.
We let $W_f^j$ denote packet $j$ from file $f \in \Fc$. We assume that each packet contains $B$ information bits, such that
$\{W_f^j\}$ are i.i.d. random variables uniformly distributed over $\{1, 2, 3, \cdots, 2^B\}$. 
We are interested in the regime of fixed $L'$ and $L \rightarrow \infty$, 
such that the probability that segments of different users overlap vanishes.  Then, have:

\begin{defn}
{\bf (Coded Cache Phase)} The caching phase is a map of the file library 
$\Fc$ onto  the cache of the users in $\Uc$.  Each cache has size $M$ files. 
For each $u \in \Uc$, the function $\phi_u: \FF_2^{mBL} \rightarrow \FF_2^{MBL}$ 
generates the cache content $Z_u \triangleq \phi_u(W_f^j, f = 1, \cdots, m,  j = 1, \cdots, L)$. 
\end{defn}

\begin{defn}
{\bf (Delivery Phase)} Let $R_u^{\rm T}$ denote the number of bits needed to be transmitted by 
node $u$ to satisfy the request vector $\fsf$. 
Then, we define the rate of node $u$ as $R_u = \frac{R_u^{\rm T}}{BL'}$.  
The function $\psi_{u}: \FF_2^{MBL} \times \Fc^n \rightarrow \FF_2^{BL'R_u}$ generates the transmitted message 
$X_{u,\fsf}  \triangleq \psi_{u}(Z_u,\fsf)$ of node $u$ as a function of its cache content $Z_u$ and of the demand vector $\fsf$. 
We denote the set of nodes whose transmitted information is useful at node $u$ is $\mathcal{D}_u$. 
The function $\lambda_u : \FF_2^{BL'\sum_{i \in \mathcal{D}_u}R_i} \times \FF_2^{MBL} \times \Fc^n \rightarrow \FF_2^{BL'}$
decodes the request of user $u$ from all messages received by users $\Dc_u$ and its own cache, i.e., we have
\be
\hat{W}_{u,\fsf} \triangleq \lambda_u(\{ X_{i,\fsf} : i \in \Dc_u\}, Z_u, \fsf). 
\ee
\hfill $\lozenge$
\end{defn}

The worst-case error probability is defined as
\be
P_e = \max_{\fsf \in \Fc^n} \max_{u \in \Uc} \PP\left( \hat{W}_{u,\fsf}  \neq (W_{f_u}^{s_u}, \ldots, W_{f_u}^{s_u+L'-1}) \right).
\ee
Letting $R = \sum_{u \in \Uc} R_u$, the cache-rate pair $(M, R)$ is achievable if $\forall$ $\varepsilon > 0$ 
there exist a set of cache encoding functions $\{\phi_u\}$, 
a set of delivery functions $\{\psi_u\}$ and a set of decoding functions $\{\lambda_u\}$ 
such that $P_e < \varepsilon$. Then the optimal achievable rate~\footnote{As a matter of fact, this is the min-max number of packet transmissions where 
min is over the caching/delivery scheme and max is over the demand vectors, and thus intuitively is the inverse of the "rate" commonly used in communications theory. We use the term ``rate'' in order to stay compliant with the terminology introduced in 
\cite{maddah2012fundamental}.} is given by 
\be
R^*(M) \triangleq \inf\{R : (M, R) \text{ is achievable}\}.
\ee
In order to relate the rate to the throughput of the network, defined later, we introduce the concept of transmission policy. 

\begin{defn} 
{\bf (Transmission policy)} The transmission policy $\Pi_t$ is a rule to activate 
the D2D links in the network. Let $\Lc$ denote the set of all directed links.  
Let $\Ac \subseteq 2^\Lc$ the set of all possible feasible subsets of links 
(this is a subset of the power set of $\Lc$, formed by all sets of links forming independent sets in the 
network interference graph induced by the protocol model).  
Let $\Asf \subset \Ac$ denote a feasible set of simultaneously active links. 
Then, $\Pi_t$ is a conditional probability mass function over $\Ac$ given $\fsf$ (requests) and the coded caching functions,
assigning probability $\Pi_t(\Asf)$ to $\Asf \in \Ac$. 
\hfill $\lozenge$
\end{defn}

In this work, we use a deterministic transmission policy, which is a special case of random policy defined above. 
Suppose that for a given caching/delivery scheme $(M,R)$ is achievable. Suppose also that 
for a given transmission policy $\Pi_t$, the $R BL'$ coded bits to satisfy the worst-case demand vector
can be delivered in $t_s$ 
channel uses (i.e., it takes $t_s$ channel uses to deliver the required $BL' R_u$ coded bits
to each user $u \in \Uc$, where each channel use 
carries $C_r$ bits). Then, the throughput per user 
measured in useful information bits per channel use 
is given by  
\begin{align}  
\label{useful-throughput-i}
T &\triangleq \frac{BL'}{t_s}. 
\end{align}

The pair $(M, T)$  is achievable if  $(M,R)$ is achievable, 
and if there exists a transmission policy $\Pi_t$ such that the $RBL'$ encoded bits can be delivered to their destinations
in $t_s \leq (B L')/T$ channel uses. 
Then, the optimal achievable throughput is defined as
\be
T^*(M) \triangleq \sup\{T : (M, T) \text{ is achievable}\}
\ee
In the following we assume that  the necessary condition $Mn \geq m$ such that any demand can be satisfied. Otherwise, 
the file library cannot be entirely cached in the union of the user caches, and some demands cannot be satisfied. 
With random demands, such setting can be handled by defining a throughput versus outage probability tradeoff, as we did in \cite{ji2013optimal}.
However, random demands are not considered in this work. 

We observe that our problem includes two parts: 1) the design of the caching, delivery and decoding functions; 2) scheduling concurrent transmissions in the D2D network. 
In the analysis, for simplicity, we start by not considering the scheduling problem and let the transmission range $r$ such that any node can be heard by all other nodes 
($r \geq \sqrt{2}$). In this case, only one simultaneous active link can be supported by the network. Then, we will relax the constraint on the transmission range 
$r$ and consider spatial reuse and scheduling.

\section{Main Results}
\label{sec: Main Results}

\subsection{Transmission range $r \geq \sqrt{2}$}

For $r \geq \sqrt{2}$, the actual users spatial distribution is irrelevant. 
The following theorem yields the achievable rate obtained by our proposed constructive coded caching and delivery
scheme.
\begin{theorem}
\label{theorem: 1}
For $r \geq \sqrt{2}$ and $t = \frac{Mn}{m} \in \mathbb{Z}^+$, the following rate is achievable:
\be
\label{eq: theorem 1}
R(M) = \frac{m}{M}\left(1-\frac{M}{m}\right).
\ee
Moreover, when $t$ is not an integer, the convex lower envelope of $R(M)$ is achievable.
\hfill  $\square$
\end{theorem}

The caching and delivery scheme achieving (\ref{eq: theorem 1}) is given in  \cite{mingyue20132} and an illustrative example is given in 
Section \ref{sec:example}.
The corresponding achievable throughput is given by the following immediate corollary.
\begin{corollary}
\label{corollary: 1}
For $r \geq \sqrt{2}$,  the throughput 
\be  \label{cor 1}
T(M) = \frac{C_r}{R(M)},
\ee
where $R(M)$ is given by (\ref{eq: theorem 1}) is achievable.
\hfill  $\square$
\end{corollary}
\begin{IEEEproof} In order to deliver $BL'R(M)$ coded bits without reuse (at most one active link transmitting at any time)
we need $t_s = BL'R(M)/C_r$ channel uses. Therefore, (\ref{cor 1}) follows from the definition (\ref{useful-throughput-i}). 
\end{IEEEproof}
A lower bound (converse result) for the achievable rate in this case is given by the following theorem.
\begin{theorem}
\label{theorem: 2}
For $r \geq \sqrt{2}$, any achievable rate is lower bounded by 
\begin{align}
R^*(M) \geq & \max\left\{\max_{s \in \{1, 2, \cdots, \min\{m, n\}\}} \left(s - \frac{s}{\lfloor\frac{m}{s}\rfloor}M\right),  \right. \notag\\
& \left. \frac{n}{n-1}\left(1-\frac{M}{m}\right)\right\}.  \label{banana}
\end{align}
\hfill  $\square$
\end{theorem}
Given the fact that activating a single link per channel use is the best possible feasible
transmission policy, we obtain trivially that using the lower bound (\ref{banana}) {\em in lieu} of $R(M)$ in 
(\ref{cor 1}) we obtain an upper bound to any achievable throughput. 
The order optimality of our achievable rate is shown by:
\begin{corollary}
\label{corollary: 3}
When $t = \frac{Mn}{m} \in \mathbb{Z}^+$, the ratio of the achievable over the optimal rate is upper bounded by 
\be
\label{eq: corollary 3}
\frac{R(M)}{R^*(M)} \leq \frac{1}{M(1-A)\left(1 - \frac{(1-A)M}{A}\right)},
\ee
where $A = \sqrt{1-\frac{1}{M+1}}$. 
\hfill  $\square$
\end{corollary}
Obviously, the same quantity upper bounds the ratio $\frac{T^*(M)}{T(M)}$.

\subsection{Transmission range $r < \sqrt{2}$}

In this case, the transmission range can be picked arbitrarily in order to force D2D communication to be localized and allow for some spatial reuse.
In this case, we need to design also a transmission policy to schedule concurrent active links. The proposed policy is based on {\em clustering}:
the network is divided into clusters of equal size $g_c$, independently of the users' demands. 
Users can obtain the requested files only from nodes in the same cluster. 
Therefore, each cluster is treated as a small network. Assuming that $g_c M \geq m$, the total storage capacity of each cluster is sufficient to store the whole
file library. Under this assumption, the same caching and delivery scheme used to prove Theorem \ref{theorem: 1}
can be used here.  A simple achievable transmission policy consists of partitioning the set of clusters into
$K$ subsets, such that the clusters of the same set do not interfere, activate simultaneously one link per cluster in each subset, and use TDMA in order to
avoid interference between the clusters. This is a classical time-frequency reuse scheme with reuse factor $K$ \cite[Ch. 17]{molisch2011wireless}, 
as shown in Fig.~\ref{fig: Grid_TDMA}. In particular, we can pick $K = \left(\left\lceil\sqrt{2}(1+\Delta)\right\rceil+1\right)^2$. 
This scheme achieves the following throughput:

\begin{theorem}
\label{theorem: 3}
Let $r$ such that any two nodes in a ``squarelet'' cluster of size $g_c$ can communicate, and  
let $t = \frac{g_cM}{m} \in \mathbb{Z}^+$. Then, the throughput
\be
T(M) = \frac{C_r}{K} \frac{1}{R(M)},
\ee
is achievable,  where $R(M)$ is given by \emph{Theorem}~\ref{theorem: 1}, 
$r$ is the transmission range and $K$ is the TDMA parameter. 
Moreover, when $t \notin \mathbb{Z}^+$, $T(M)$ can be computed by using the convex lower envelope of $R(M)$.
\hfill  $\square$
\end{theorem}
Notice that whether reuse is convenient or not in this context depends on whether 
$C_{\sqrt{2}}$ (the link spectral efficiency for communicating across the network) is larger or smaller than $C_r/K$, for some smaller $r$ which determines the 
cluster size. In turns, this depends on the dependency of the link spectral efficiency  on the communication range. This aspect is not captured by the protocol model, 
and the answer may depend on the operating frequency and appropriate channel model of the underlying wireless network physical layer.  

An upper bound on the throughput with reuse is given by:
\begin{theorem}
\label{theorem: 4}
When $r < \sqrt{2}$ and the whole library is cached within radius $r$ of any node, 
the throughput is upper bounded by 
\be
T^*(M) \leq \frac{C_r \left\lceil\frac{64}{\Delta^2} \right\rceil}{\max_{s \in \{1, 2, \cdots, \min\{m, \lceil \pi r^2 n\rceil \}\}} \left(s - \frac{s}{\lfloor\frac{m}{s}\rfloor}M\right)},
\ee
where $r$ is the transmission range and $\Delta$ is the interference parameter. 
\hfill  $\square$
\end{theorem}

Similarly to the case of $r \geq \sqrt{2}$, we have the upper bound on the optimal to achievable throughput ratio:
\be
\frac{T^*(M)}{T(M)} \leq \frac{K \left\lceil\frac{64}{\Delta^2} \right\rceil  }{M(1-A)\left(1 - \frac{(1-A)M}{A}\right)},
\ee
where $A = \sqrt{1-\frac{1}{M+1}}$ and for $t = \frac{Mg_c}{m} \in \mathbb{Z}^+$. 

\section{An Example}  \label{sec:example}

\begin{figure}
\centering
\includegraphics[width=7cm]{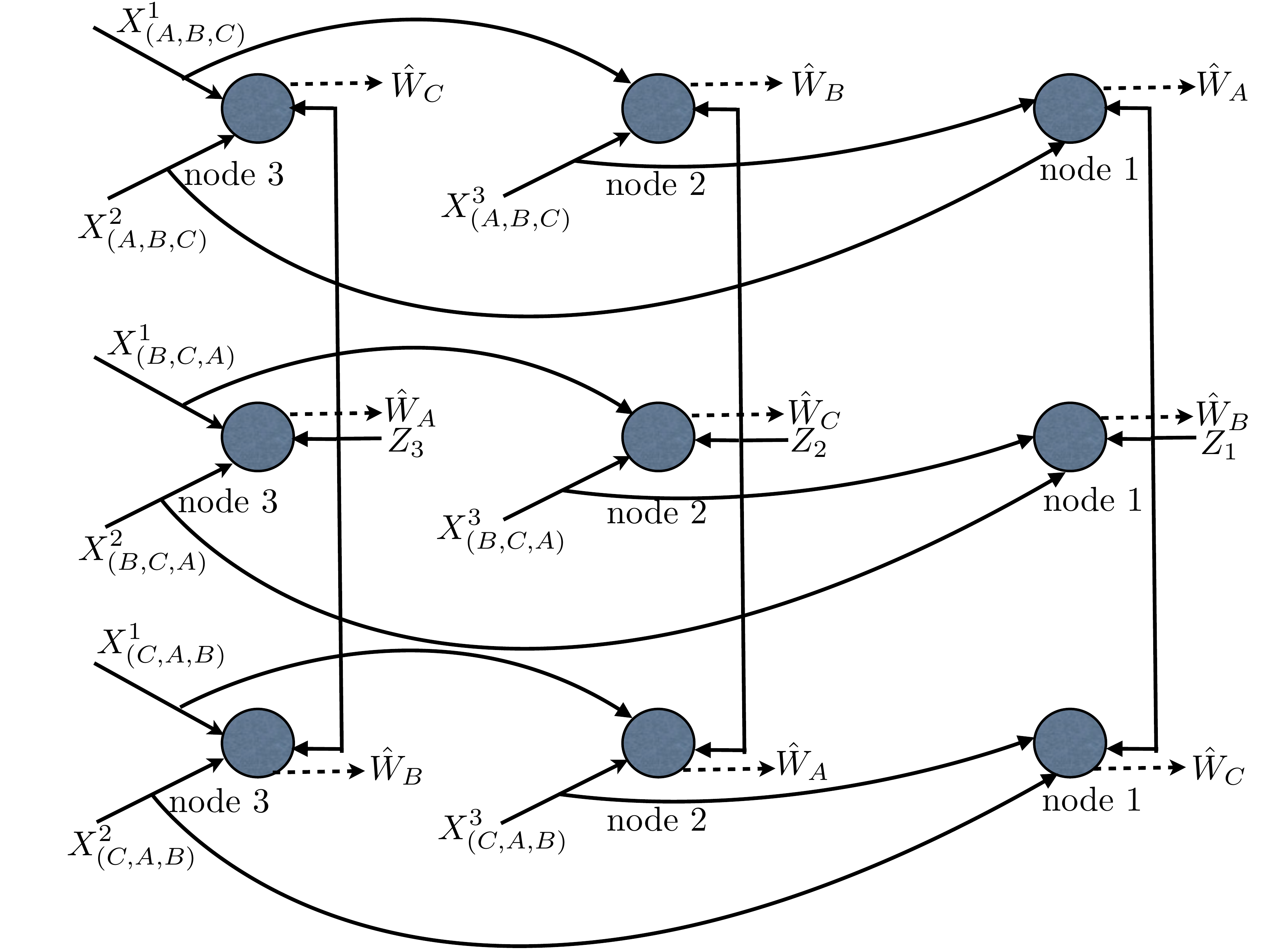} 
\caption{The augmented network when $m=3$, $n=3$. The three requested vectors are: $(A,B,C)$, $(B,C,A)$ and $(C,A,B)$.}
\label{fig: Network_Coding_Model}
\end{figure}

The proposed caching placement and delivery scheme is illustrated through a simple example.
Consider a network with three users ($n=3$). 
Each user can store $M=2$ files, and the library has size $m=3$ files, 
which are denoted by $A, B, C$.  Let $r \geq \sqrt{2}$.
Without loss of generality, we assume that each node requests one packet of a file ($L' = 1$).
We divide each packet of one file into $6$ sub-packets, and  denote the sub-packets of the $j$-th 
packet as 
$\{A_{j,\ell} : \ell = 1, \ldots, 6\}$,
$\{B_{j,\ell} : \ell = 1, \ldots, 6\}$, and
$\{C_{j,\ell} : \ell = 1, \ldots, 6\}$.
The size of each sub-packet is $F/6$. 
We let user $u$ stores $Z_u$, $u = 1, 2, 3$, given as follows:
\begin{align}
Z_1 = &(A_{j,1}, A_{j,2}, A_{j,3}, A_{j,4}, B_{j,1}, B_{j,2}, B_{j,3}, B_{j,4}, \notag\\
&C_{j,1}, C_{j,2}, C_{j,3}, C_{j,4}),  j = 1, \cdots, L.
\end{align}
\begin{align}
Z_2 = &(A_{j,1}, A_{j,2}, A_{j,5}, A_{j,6}, B_{j,1}, B_{j,2}, B_{j,5}, B_{j,6}, \notag\\
&C_{j,1}, C_{j,2}, C_{j,5}, C_{j,6}), j = 1, \cdots, L.
\end{align}
\begin{align}
Z_3 = &(A_{j,3}, A_{j,4}, A_{j,5}, A_{j,6}, B_{j,3}, B_{j,4}, B_{j,5}, B_{j,6}, \notag\\
&C_{j,3}, C_{j,4}, C_{j,5}, C_{j,6}), j = 1, \cdots, L.
\end{align}
In this example, we consider the demand $\fsf = (A, B, C)$ with initial point in the requested segment $\ssf = (1,2,3)$, i.e., 
user $1$ requests packet $1$ of file $A$, user $2$ requests packet $2$ of file $B$ and user $3$ requests packet $3$ of file $C$. 
Then, the delivery scheme is the following. User $1$ transmits $B_{2, 3} + C_{3, 1}$. 
User $2$ transmits $A_{1,5} + C_{3,4}$. User $3$ transmits $A_{1,6}+B_{2,4}$. 
Thus, $R_1+R_2+R_3 = \frac{1}{6} \cdot 3 = \frac{1}{2}$.


Next, we illustrate the idea of the general achievable rate lower bound of Theorem \ref{theorem: 2}.
Without loss of generality, we assume that $L/L'$ is an integer and let $s$ denote the segment index.
For any scheme that satisfies arbitrary demands $\fsf$, with arbitrary segments $\ssf$, we denote by 
$R_{u,s,\fsf}^{\rm T}$ the number of transmitted bits for user $u$, requesting segment $s$ when the request vector is $\fsf$.  
Since the requests are arbitrary, we can consider a time extension for all possible request vectors. 
For example, we let the first request be $\fsf=(A,B,C)$, the second request be $\fsf = (B, C, A)$ and the third request be $\fsf = (C,A,B)$.
Then, the augmented time-extended graph is shown in  Fig.~\ref{fig: Network_Coding_Model}. 
Considering user 3, from the cut that separates $(X_{1,(A,B,C)}, X_{2,(A,B,C)}, X_{1,(B,C,A)}, X_{2,(B,C,A)}, X_{1,(C,A,B)},$\\ 
$X_{2,(C,A,B)}, Z_3)$ and $(\hat{W}_C, \hat{W}_A, \hat{W}_B)$, we can obtain that
\begin{align} \label{eq: converse 1}
&\sum_{s=1}^{\frac{L}{L'}} \left(R_{1,s,(A,B,C)}^{\rm T}+R_{2,s,(A,B,C)}^{\rm T} + R_{1,s,(B,C,A)}^{\rm T} +R_{2,s,(B,C,A)}^{\rm T} \right. \notag\\
&\left.  + R_{1,s,(C,A,B)}^{\rm T}+R_{2,s,(C,A,B)}^{\rm T} \right) + MBL \geq 3BL' \cdot L/L.
\end{align}
Similarly, from the cut that separates $(X_{1,(A,B,C)}$, $X_{3,(A,B,C)}$, $X_{1,(B,C,A)}$, $X_{3,(B,C,A)}$, $X_{1,(C,A,B)}$,
$X_{3,(C,A,B)}$, $Z_2)$ and $(\hat{W}_B, \hat{W}_C, \hat{W}_A)$, 
and from the cut that separates $(X_{2,(A,B,C)}$, $X_{3,(A,B,C)}$, $X_{2,(B,C,A)}$, $X_{3,(B,C,A)}$, $X_{2,(C,A,B)}$,
$X_{3,(C,A,B)}$, $Z_1)$ and $(\hat{W}_A, \hat{W}_B, \hat{W}_C)$, 
we can obtain similar formulas. By summing (\ref{eq: converse 1}) and the other two corresponding formulas and dividing all terms by 2, we obtain 
\begin{align}
&\sum_{s=1}^{\frac{L}{L'}}  \left(R_{1,s,(A,B,C)}^{\rm T} + R_{2,s,(A,B,C)}^{\rm T} + R_{3,s,(A,B,C)}^{\rm T} + R_{1,s,(B,C,A)}^{\rm T} \right. \notag\\
&\left. + R_{2,s,(B,C,A)}^{\rm T} + R_{3,s,(B,C,A)}^{\rm T} + R_{1,s,(C,A,B)}^{\rm T} + R_{2,s,(C,A,B)}^{\rm T} \right. \notag\\
&\left.  + R_{3,s,(C,A,B)}^{\rm T} \right) + \frac{3}{2}MBL \geq \frac{9}{2} BL.
\end{align}
Noticing that, by symmetry, $R^{\rm T}  = R^T_{1,s,\fsf} + R^T_{2,s,\fsf} + R^T_{2,s,\fsf}$ for any $\ssf$ and $\fsf$, we have
\be
\frac{3 L}{L'} R^{\rm T} \geq \frac{9}{2} BL  - \frac{3}{2}MBL.
\ee
Dividing both sides by $3BL$, we obtain that any achievable coding scheme must satisfy
\be
R(M) = \frac{R^{\rm T}}{BL'} \geq \frac{3}{2} - \frac{1}{2}M.
\ee
In the example of this section, for $M = 2$ we obtain $R^*(2) \geq \frac{1}{2}$. Therefore, in this case the achievability scheme is optimal. 

Considering the same network $n=3$ users with storage capacity $M=2$ and library size $m = 3$, but adding a special node (base station) 
with all files available, the coded caching scheme of  \cite{maddah2012fundamental} achieves $R(2) = \frac{1}{3}$. 
Then, in this case, the relative loss incurred by not having a base station with all files available is $3/2$. 

\section{Discussions}
\label{sec: Discussions}

The achievable rate of  \emph{Theorem}~\ref{theorem: 1} can be written as the product of three terms, 
$R(M) = n\left(1-\frac{M}{m}\right)\frac{m}{Mn}$ with the following interpretation: 
$n$ is the number of transmissions by using a conventional scheme that serves individual demands without exploiting the demand redundancy; 
$\left(1-\frac{M}{m}\right)$ can be viewed as the {\em local caching gain}, any user can cache a fraction $M/m$ of any file, therefore it needs to receive
only the remaining part;  $\frac{m}{Mn}$ is the {\em global caching gain}, i.e., the ability of coded caching and delivery to turn the individual but overlapping demands into
a coded multicast, such that transmissions are useful to many users despite the streaming sessions are strongly asynchronous. These three terms with the same interpretation
can be found also in the rate expression of the scheme in \cite{maddah2012fundamental} (see Section \ref{section: intro}), 
where the base station has access to  all the files. Comparing this rate with our Theorem~\ref{theorem: 1}, we notice that they differ only in the last term
(global caching gain), which in the base station case is given by $(1+\frac{nM}{m})^{-1}$. For $nM \gg m$, we notice that these factors are essentially identical. 

The lower bound gap of (\ref{eq: corollary 3}) shows that, when $M$ is a constant, and $m \gg 1$, 
the achievable rate achieves the same order of the converse. The multiplicative gap between the achievable rate and the converse lower bound 
is a decreasing function of $M$ between $5.83$ (for $M = 1$), and $4$ (for $M$ asymptotically large). 

As already noticed, \emph{Theorem}~\ref{theorem: 3} shows that there is no fundamental cumulative gain by using both 
spatial reuse and coded caching. Under our assumptios, spatial reuse may or may not be convenient depending whether
$\frac{C_r}{K}$ is larger or smaller than $C_{\sqrt{2}}$. A closer look reveals a more subtle tradeoff. 
Without any spatial reuse, the length of the codewords for each user, related to the size of the sub-packetization, 
is ${n \choose \frac{Mn}{m}}$. This may be very large when $n$ and $M$ are large. At the other extreme, we have the case where the cluster size is 
the minimum able to cache the whole library in each cluster. In this case, we can just store $M$ different whole files into each node, such that all $m$ files are present in each cluster, and for the delivery phase we just serve whole files without any coding as in \cite{ji2013optimal}. 
In this case, the achieved throughput is $\frac{C_r}{K} \frac{M}{m}$ bits/s/Hz, 
which is almost as good as our proposed scheme ($\frac{C_r}{K\left(\frac{m}{M}-1\right)}$). 
This simple scheme is a  special case of the general setting treated in this paper, where 
spatial reuse is maximized and codewords have length 1. If we wish to use the achievable scheme of this paper, 
the codewords length is ${g_c \choose \frac{M g_c}{m}}$. 
Hence, spatial reuse yields a reduction in the codeword length of the corresponding coded caching scheme.

\bibliographystyle{IEEEbib}
\bibliography{references}

\end{document}